\documentclass[12pt]{article}
\usepackage{epsfig}
\begin{document}
\begin{center}
{\bf \Large Threshold meson production in nucleon-nucleon collisions\footnote{
Contribution to "MESON96", Workshop on production, properties and interaction of mesons,
 Cracow, Poland, 10-14 May 1996.}}
\end{center}
\begin{center}
  J. Haidenbauer, Ch. Hanhart and J. Speth\\
{\it Institut f\"{u}r Kernphysik, Forschungszentrum J\"{u}lich
GmbH, D--52425 J\"{u}lich, Germany}
\end{center}

\vspace{8mm}

\begin{center}
{\bf Abstract}
\end{center}
\begin{small}
A brief overview on recent model calculations for near threshold
pion production in nucleon-nucleon collisions is given. Results from
our own investigations of the reactions $pp
\rightarrow pp\pi ^0$ and $pn \rightarrow d\pi^0$ are presented.
Direct production, heavy meson exchange and pion
rescattering are taken into account. For the latter a T-matrix
obtained from a microscopic model of the $\pi N$ interaction is
employed. We obtain a significant
contribution from rescattering, but not enough to describe the
data for $pp \to pp\pi^0$. The missing production rate can be
provided by heavy meson exchanges.
For the first time the effect of off--shell rescattering
is investigated for the reaction $pn \rightarrow d\pi^0$.
Isoscalar rescattering in combination with isovector
rescattering is able to describe the s-wave production data.
We confirm that heavy meson exchanges are negligible in this
reaction.
\end{small}

\vspace{8mm}

\section{ Introduction}

Recent advances in accelerator technology have opened new perspectives
in nuclear physics \cite{Pol}. The possibility of doing experiments
with internal targets in storage rings together with the tool of
beam cooling have made it feasible to study particle production
processes extremely close to their thresholds and with
unprecedented accuracy. Such data have been eagerly awaited by
theorists. In the proximity of the threshold the production
processes are determined by only a few amplitudes and therefore
a theoretical interpretation of them should be simple but at the
same time also rather informative.

\begin{table}[h]
\label{tab}
\begin{center}
\begin{tabular}{|c||c|c|}
\hline
& & \\
& {\large $pp \to pp\pi^0$} & {\large $pn \to d\pi^0$} \\
& & \\
& $|f> = |^1S_0 l_\pi >$ & $|f> = |(^3S_1-^3D_1) l_\pi >$ \\
& & \\
\hline
& & \\
$l_\pi=0$  & $|i>=|^3P_0>$ &  $|i>=|^3P_1>$ \\
& & \\
\hline
& & \\
$l_\pi=1$  & -- &  $|i>=|^1S_0>,|^1D_2>$ \\
& & \\
\hline
\end{tabular}
\end{center}

\caption{Allowed partial waves near threshold for the reactions
$pp \rightarrow pp\pi^0$ and $pn \rightarrow d\pi^0$. The notation
$^{2S+1}L_J$ is used for the $NN$ states. $l_\pi$ is the angular
momentum of the pion relative to the $NN$ system. $|i>$ and $|f>$
denote the initial and final states, respectively.}
\end{table}

The reaction $NN \rightarrow NN\pi$ is of particular interest
because it constitutes the dominant inelastic process in the
$NN$ interaction. For the pion production at threshold a large
momentum transfer of typically 370 MeV/c between the nucleons is
required. This corresponds to $NN$ separations of roughly 0.5 fm.
Consequently the study of the pion production process can provide
us with informations about the short-range part of the $NN$ interaction.
Since the pion can rescatter on the nucleon before it is emitted it
could be also possible to learn something about the off-shell properties
of the $\pi N$ interaction. Furthermore  the reaction $NN \rightarrow NN\pi$
can serve as a testing ground for theoretical models of meson
production which can then be applied to the production of
other, heavier mesons such as $\eta$, $\eta '$ or $\phi$.

Over the last few years several near-threshold experiments for
various charge channels of the reaction $NN \rightarrow NN\pi$ were
performed. The reaction $pp \rightarrow pp\pi^0$ was
measured at the Cyclotron Facility of the Indiana University
(IUCF) \cite{IU1,data}, at TRIUMF \cite{TR1}, and more recently
also in Uppsala \cite{UP}.
Data on $pp \rightarrow d\pi^+$ were provided by the
TRIUMF group \cite{TR2}, by the GEM collaboration at COSY in
J\"ulich \cite{CO} and by IUCF \cite{IU3}. Finally, there are
measurements on the reactions $pn \rightarrow d\pi^0$ \cite{Hu},
$pn \rightarrow pp\pi^-$ \cite{Ba}, and
$pp \rightarrow pn\pi^+$ \cite{Da}. Many of the data are taken
at values of $\eta$ (the maximum center-of-mass momentum of the
produced pion divided by the pion mass, $q^\pi_{max}/m_\pi$) below 0.5.
This corresponds to bombarding energies of less than 50 MeV
above threshold.

At such energies very close to threshold the angular momentum
$L_{NN}$ in the final
$NN$ system as well as the angular momentum $l_{\pi (NN)}$ of the
pion with respect to the $NN$ pair is restricted to the values
0 and 1. Conservation laws for the angular momentum, isospin,
and parity, and the Pauli principle limit the number of possible
partial wave amplitudes further. The allowed partial waves under the
restrictions $L_{NN} = 0$ and $l_\pi = 0,1$ are summarized in Table 1
for the reactions $pp \rightarrow pp\pi^0$ and
$pn \rightarrow d\pi^0$.
These partial waves will give the dominant contributions to the
pion production at threshold. $NN$ states with $L_{NN}=1$ are
much less important since the interaction in these partial waves
is comparatively weak. Note that
there is no entry for the $l_\pi = 1$ contribution
in case of $pp \rightarrow pp\pi^0$. This means that (p-wave)
pion production via the $\Delta$ (1236) resonance,
which plays a major role in other meson production reactions, is
strongly suppressed in this particular channel.

\section{ Models for threshold pion production: A brief history}

\begin{figure}[h]
\begin{center}
\epsfig{file=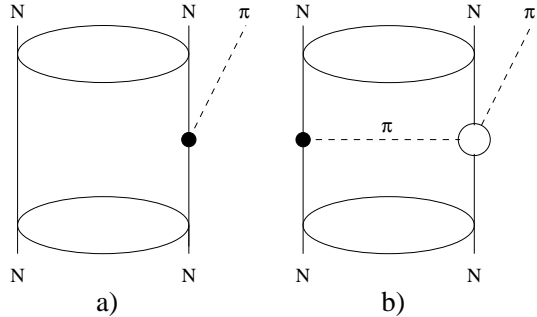, height=7cm, angle=-90}
\caption{Pion production mechanisms included in the model by Koltun
and Reitan: (a) direct pion emission, (b) pion rescattering.
}
\label{pm1}
\end{center}
\end{figure}

Essentially all recent theoretical investigations on pion production
near threshold build up on the model proposed by
Koltun and Reitan in 1966 \cite{KuR}. In this model two production
mechanisms are considered: (i) direct production of a pion,
depicted in Fig.~\ref{pm1}a, (ii) production of a pion which first
scatters off the other nucleon before emission, as shown in
Fig.~\ref{pm1}b. The former is usually called Born term
whereas the latter is referred to as rescattering term.
This model was utilized by Miller and Sauer in 1991 \cite{MuS}
to analyze the first set of high precision data of the
reaction $pp \rightarrow pp\pi^0$ near threshold that became available
from IUCF \cite{IU1}.
Surprisingly, it turned out that such a model grossly underestimates
the empirical cross section \cite{MuS}. Only the predicted energy
dependence of the cross section, which in this model is determined
essentially by phase-space factors and the $pp$ final-state
interaction, was found to be in agreement with the IUCF data \cite{data}.

We want to remark that in this model the $\pi N$ interaction
occurring in the rescattering diagram is approximated by the $\pi N$
(s-wave) scattering length. In the reaction $pp \rightarrow pp\pi^0 $
only the isoscalar component of the $\pi N$ s-wave interaction is
present. Since the corresponding scattering length is almost zero due
to chiral constraints, it means that the contribution from the
rescattering process is practically negligible \cite{MuS}.

\begin{figure}[h]
\begin{center}
\epsfig{file=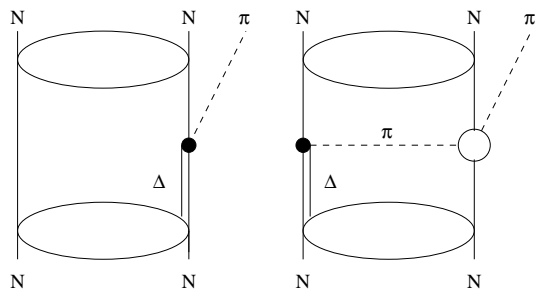, height=7cm, angle=-90}
\caption{Pion production via $\Delta$ excitation.}
\label{pm2}
\end{center}
\end{figure}

In 1992 Niskanen extended this model by including in
addition pion rescattering in the $\pi N$ p-wave via the $\Delta$ (1236)
isobar (cf. Fig.~\ref{pm2}) \cite{Nis1}. Furthermore, he allowed for
an energy dependence in the s-wave rescattering term \cite{Nis2}. These
improvements roughly doubled the predicted $pp\rightarrow pp\pi^0$
cross section, but Niskanen's results still underestimate the IUCF data
by a factor of 3.6.

\begin{figure}[h]
\begin{center}
\epsfig{file=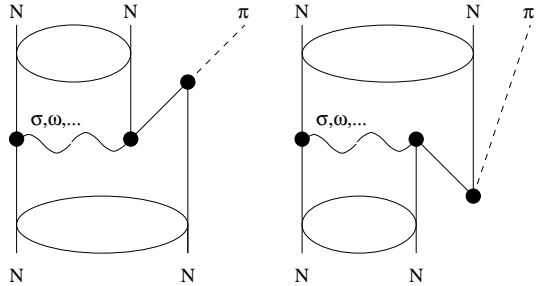, height=7cm, angle=-90}
\caption{Mechanisms for pion production:
Heavy meson exchanges.
}
\label{pm3}
\end{center}
\end{figure}

Another new production mechanism was introduced by Lee and Riska
in 1993 \cite{LuR}. These authors considered effects from meson-exchange
currents due to the exchange of heavy mesons, as shown in Fig.~\ref{pm3}.
It was found that the resulting contributions (in particular the
one of the $\sigma$ meson) enhance the pion production cross section
by a factor of 3-5 \cite{LuR,HMG} and thus eliminate most of the
underprediction found in earlier investigations.
Consequently, at that time it seemed that the reaction $pp\rightarrow
pp\pi^0$ near threshold is essentially understood.

However, in 1995 Hern\'andez and Oset presented an alternative
explanation for the missing strength in the $\pi^0$ production close
to threshold \cite{HO}. These authors took into account the
off-shell properties of the $\pi N$ amplitude in the
evaluation of the rescattering diagram.
Since the isoscalar s-wave $\pi N$ off-shell amplitude can be much larger
than its on-shell value at threshold (which is more or less zero,
as mentioned before) it turned out that the contribution from
rescattering is now considerably enhanced.
Indeed it was demonstrated in Ref. \cite{HO} that direct production
and rescattering alone are also sufficient in order to reproduce the
empirical $pp\rightarrow pp\pi^0$ cross section.

Yet another aspect was added to this controversial situation by
two recent investigations based on chiral perturbation theory
\cite{CP1,CP2}. In these studies a considerable cancellation
between the contributions from direct production and rescattering
is observed. Then one is essentially left with only
contributions from rather short-ranged mechanisms such as heavy
meson exchanges - which, however, are not sufficient for describing
the experiment \cite{CP2}.

\section{ The $pp \rightarrow pp\pi^0$ reaction}

In the following two sections we will report on our own investigations
of the
reaction $NN \rightarrow NN\pi$ \cite{Han1}. The main novelty in
these calculations is that a realistic meson-theoretical model of the
$\pi N$ interaction \cite{S} is employed for the evaluation of the
rescattering contributions. We present results for the channels
$pp \rightarrow pp\pi^0$ and (in the next section) $pn \rightarrow d\pi^0$.
We consider only the lowest partial waves in the outgoing channel,
i. e. the $pp$ pair is taken to be in the $^1S_0$ and the pion is in
an s-wave relative to the nucleon pair or the deuteron, respectively.
Our calculations are carried out in momentum space. Distortions in
the initial and final $NN$ states are taken into account. The
Bonn potential OBEPT \cite{Mac} is used for the $NN$ interaction.
The Coulomb interaction is included following
the method described in Ref.~\cite{Han1}.
The model is developed in the framework of time ordered perturbation
theory. Therefore it is consistent with the interactions in the
$NN$- and $\pi N$ systems which were likewise derived in time ordered
perturbation theory \cite{Mac,S}.

In our model calculation of the reaction $NN \rightarrow NN\pi$ we
consider contributions from the direct pion production,
from s-wave pion rescattering and from the heavy-meson-exchange (HME)
production mechanism.

\begin{figure}[h]
\begin{center}
\epsfig{file=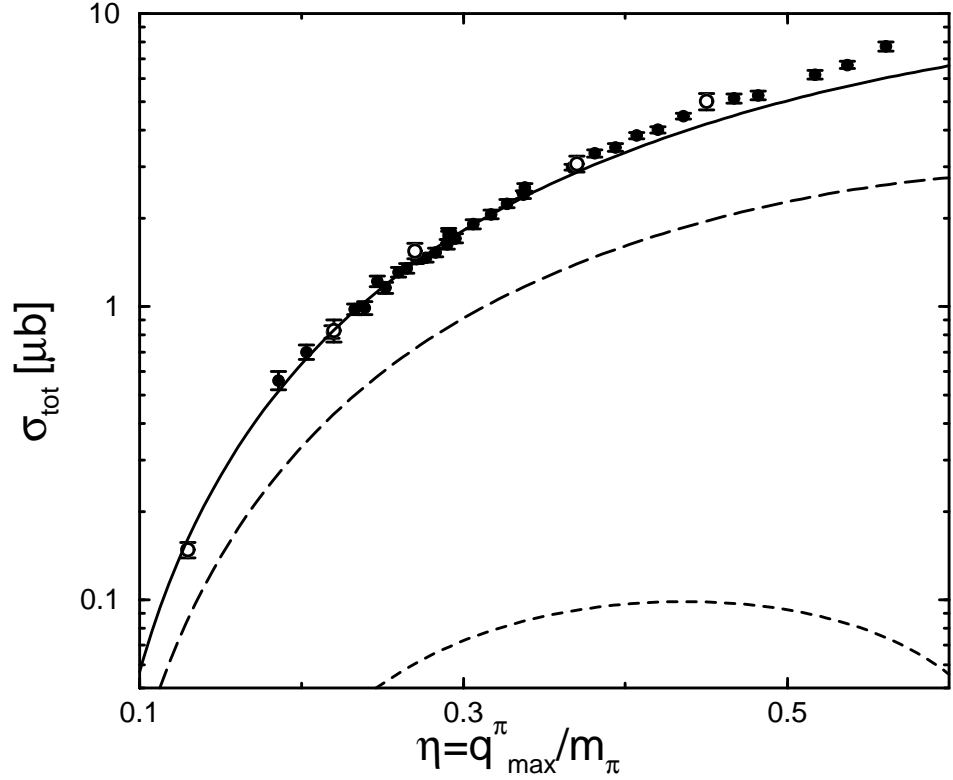, height=12cm}
\caption{Total cross section for the reaction $pp \to pp\pi^0$.
The dashed line shows the results for direct production only.
Adding rescattering yields the long dashed line. Including
also contributions from heavy meson exchanges leads to the solid
line. The data are from Refs. \protect\cite{data} (filled circles) and
\protect\cite{UP} (open circles).
}
\label{pp1}
\end{center}
\end{figure}

\subsection{The direct pion production}

Following previous investigations we use pseudo-vector coupling for
the $\pi NN$ vertex. This leads to the following structure for the
pion production vertex,
\begin{equation}
 {\sl M}_{fi} \propto
\sqrt{
 {\epsilon _p}{\epsilon _{p \, '}} \over E_p E_{p \, '}} \ \left\lbrack
\vec \sigma \cdot (\vec p- \vec p \, ')-\omega_q \vec \sigma \cdot
 \left(\frac{\vec p}{\epsilon _p}+\frac{\vec p \, '}{\epsilon _{p \, '}}\right)
\right\rbrack \ ,
\label{vertex}
\end{equation}
where $\vec p$ ($\vec p \, '$) is the incoming (outgoing) nucleon
momentum, and $\epsilon_p = E_p + M$ with the nucleon energy $E_p
=\sqrt{M^2 + \vec p \,^2 }$. $\omega_q =\sqrt{m_\pi^2 + \vec q \,^2 }$
is the energy of the pion with momentum $\vec q = \vec p - \vec p \,
'$.

In earlier calculations \cite{KuR,MuS,Nis1,LuR,HMG} several
approximations are applied in the evaluation of the production
amplitude.
The energies $E _p$, $E _{p'}$ and $\omega _q$ are replaced
by the respective masses in Eq. (\ref{vertex}) and usually the first
term in Eq. (\ref{vertex}) is omitted altogether.
Furthermore, the reduced mass of the pion relative to the $NN$ system
is replaced by the pion mass $m_\pi $ in the kinematical relations.
This increases the allowed maximum pion momentum $q^\pi_{max}$ (for a
fixed energy) and enlarges the phase space.

The consequences of these approximations were studied by us in a recent
paper\cite{Han1}. It turned out that a more correct treatment of the
direct pion production mechanism reduces its contribution
to the $pp \rightarrow pp\pi^0$ cross section by a factor of 2 and
it also modifies the energy dependence of the cross section.
In the present calculation there is a further reduction of the cross
section as compared to comparable previous investigations
which is due to the employed $NN$ interaction model. (We will comment
on the sensitivity to the $NN$ interaction later.)
Indeed the pion production cross section resulting from the Born term
alone is in our case about a factor 20 smaller than the experiment
(cf. the dashed line in figure \ref{pp1}) whereas only a factor
of about 5 is missing by, e. g., the model of Miller and Sauer \cite{MuS}.

\subsection{Pion rescattering}

\begin{figure}[h]
\begin{center}
\epsfig{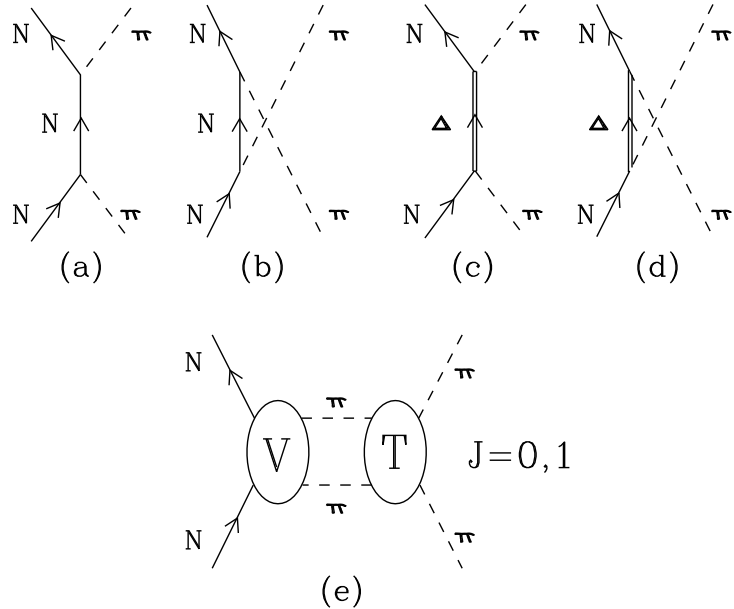}
\caption{Graphs included in the $\pi$N interaction model.
}
\label{diagr}
\end{center}
\end{figure}

The second pion production mechanism we take into account is pion
rescattering (Fig.~\ref{pm1}b).  In the model of Koltun and Reitan
the $\pi N$ scattering amplitude is derived from a phenomenological
effective Hamiltonian \cite{KuR}
\begin{equation}
 {\sl H} \ = \ 4\pi {\lambda_1 \over m_\pi} \bar \Psi \vec
 \phi \cdot \vec
\phi \Psi \ + \ 4\pi {\lambda_2 \over m_\pi^2} \bar \Psi \vec \tau
\Psi \cdot \vec \phi \times \partial _0 \vec \phi
\label{hamil}
\end{equation}
where $\lambda _1$, $\lambda _2$ are fixed by the (empirical) $S_{11}$
and $S_{31}$ pion nucleon scattering lengths. The
isovector term (proportional to $\lambda _2$) does not contribute to
the the reaction $pp \rightarrow pp\pi^0$ because of isospin constraints.
Since the isoscalar part is very small
($\lambda _1 = 0.005$ according to H\"ohler et al. \cite{hoe};
$\lambda _1 =-0.0013$ following Arndt et al. \cite{arndt}) it has
usually been found that the rescattering contribution to the $ pp
\rightarrow pp\pi ^0$ \ cross section obtained from the effective
Hamiltonian in Eq. (\ref{hamil}) is more or less negligible
\cite{MuS,HMG,LuR}.

It is well known that the smallness of the isoscalar s-wave $\pi N$
on-shell amplitude is due to a strong cancellation between different
isospin amplitudes. The situation is rather different for the
corresponding off-shell amplitude - which is actually the quantity that
enters into the evaluation of the rescattering diagram (Fig. 1b). In
order to account for these off-shell properties appropriately we
employ in our investigation a microscopic meson--exchange model of the
$\pi N$ interaction which has been constructed recently
in J\"ulich \cite{S}.  This model includes the conventional (direct and
crossed) pole diagrams involving the nucleon and the $\Delta$-isobar;
the meson exchanges in the scalar ($\sigma$) and vector ($\rho$)
channels are derived from correlated $2\pi$ exchange (Figure \ref{diagr}).
Further details about this model can be found in Ref. \cite{S}.

\begin{figure}[h]
\begin{center}
\epsfig{file=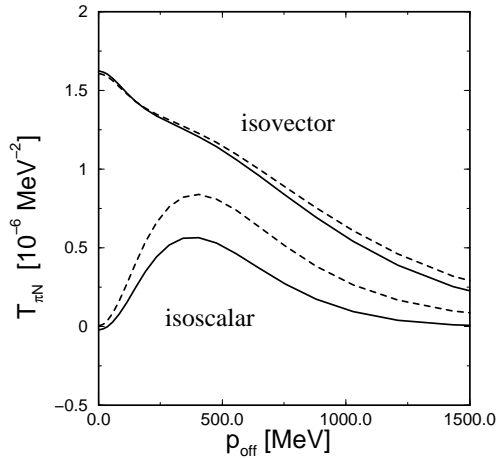, height=7cm}
\caption{Half--off--shell $\pi$N s-wave T-matrix at threshold.
The lower two curves show the isoscalar component of the T-matrix, the
upper curves display the isovector component. The solid line denotes
the model used in the present calculation whereas the
dashed line is the result from model 1 of Ref. \protect \cite{S}.
}
\label{hos}
\end{center}
\end{figure}

Here we use a slightly modified version where
the form factors are energy--independent and the antibaryon
contributions have been left out. These modifications are made to
allow an extrapolation of the model to negative energies as required
in the present three-particle context.  After readjustment of its free
parameters this model yields a good description of low--energy $\pi N$
scattering, comparable to the results shown in Ref. \cite{S}.  The
resulting $s$--wave scattering lengths are $a_1 = 0.173 m_\pi^{-1}$
and $a_3 = -0.084 m_\pi^{-1}$ leading to a value of $\lambda _1=
-0.001$ which is agreement with the value given in Ref. \cite{HMG}.
The half-off-shell T-matrix produced by this model at $\pi N$ threshold
is shown in Fig. \ref{hos}.

For the production vertex we take the same pion coupling constant
as in the $NN$ potential, namely $f^2/4\pi = 0.0795$.
The form factor is chosen to be rather soft. We use a monopole form
with a cutoff mass $\Lambda_\pi = 800 \ MeV$ which is in line with
recent QCD lattice calculation \cite{QCD} and other
informations \cite{Sca}.

Results for the $ pp \rightarrow pp\pi ^0$ \ cross section
including contributions of the rescattering mechanism are depicted
by the long-dashed curve in Fig. \ref{pp1}.
The rescattering process increases the cross section by a factor
of around 10 compared to the rate for direct production.
This is in qualitative agreement with the findings
reported in Ref. \cite{HO} which, however, are based on phenomenological
off-shell extrapolations of the $\pi N$ amplitude.

\subsection{Heavy meson exchanges}

As can be seen in Fig. \ref{pp1} our predictions based on direct
pion production plus rescattering underestimate
the empirical data \cite{data,UP} by a factor of 2.
Evidently further production mechanisms are needed.
An obvious option are corrections from meson-exchange currents as
proposed by Lee and Riska \cite{LuR}. Corresponding diagrams are
shown in Fig.~\ref{pm3}. The results presented in Ref. \cite{LuR} and the
subsequent more thorough investigations by Horowitz and collaborators
\cite{HMG} indicate that only the diagrams involving the $\sigma$
and $\omega$ mesons give rise to an appreciable contribution. Therefore
we restrict our calculation to these two mesons. The vertex parameters
for the $\omega$ exchange are taken over from OBEPT
($g^2_{\omega}/4\pi$ = 20, $\Lambda_\omega = 2000 \ MeV$).
Since the $\sigma$ meson that is used in one-boson-exchange
models of the $NN$ interaction is an effective
parameterization of more complex processes like correlated and
uncorrelated $\pi \pi$-exchange \cite{Kim} it should be
different in the present context involving vertices with antinucleons.
Hence we consider the $\sigma$ coupling constant as a free parameter
which is chosen to reproduce the $\pi^0$ production cross section close
to threshold. With the values $g^2_{\sigma}/4\pi$ = 5.7, $\Lambda_\sigma
= 1700 \ MeV$ the solid line in Fig.\ref{pp1} is obtained.
It is interesting that the required $\sigma$ coupling strength is
almost identical to the one used in the full Bonn $NN$ model
\cite{Mac}.

\subsection{Discussion}

Obviously the reaction $pp \rightarrow pp\pi^0$ is rather
sensitive to the $\pi N$ off-shell behavior as well as to the
short range component of the $NN$ force. However, in order to
learn something about either of these features it is necessary to
have some constraints on one of them. With regard to the properties
of the $\pi N$ amplitude this seems to be difficult at present.
For example, the $\pi N$ model used in the present calculation
as well as the initial model (model 1 of Ref. \cite{S}) yield an equally good
fit of the relevant $\pi N$ phase shifts.
However, due to minor differences in their dynamical ingredients
(antibaryon contributions are left out in the model applied here), the
isoscalar
s-wave $\pi N$ half-off-shell amplitude at threshold is about 50 \%
larger around the maximum in the initial model (cf. Fig.~\ref{hos})
and accordingly an enhanced contribution from the rescattering
mechanism can be expected.
Note that the isovector s-wave $\pi N$ amplitude, which is also shown in
Fig.~\ref{hos}, is much less model dependent.

\begin{figure}[h]
\begin{center}
\epsfig{file=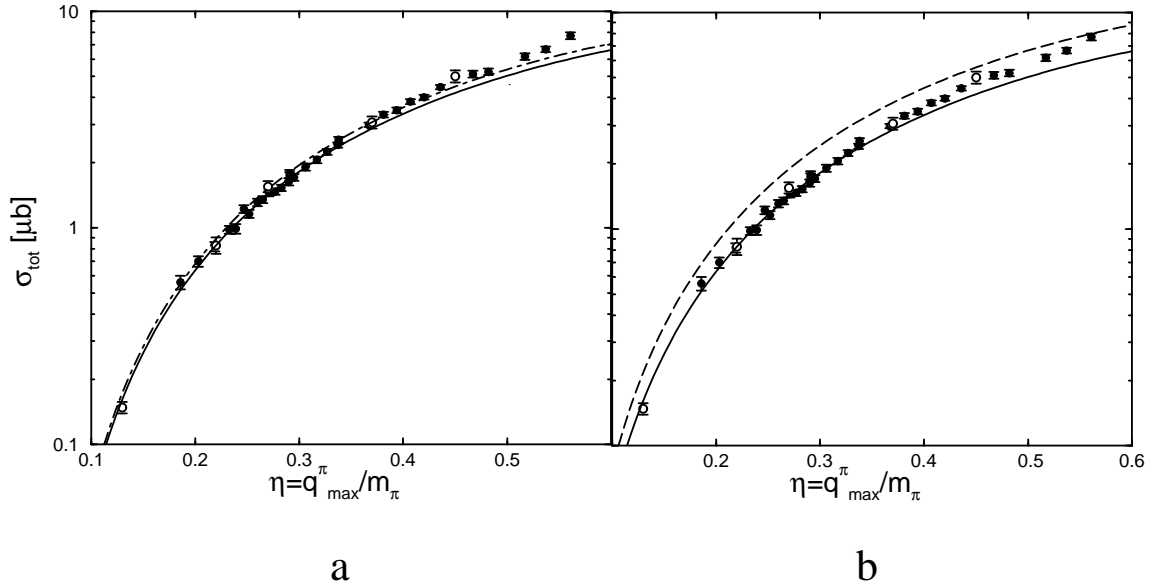, height=8cm}
\caption{Sensitivity of the reaction $pp \to pp\pi^0$ to (a) the
$\pi$NN form factor and (b) different nucleon--nucleon interactions.
The solid line is obtained with OBEPT and $\Lambda_\pi$ = 800 MeV.
The dashed--dotted line (in (a)) corresponds to $\Lambda_\pi$ =
1000 MeV. The dashed line (in (b)) is the result for the Paris
$NN$ potential.
}
\label{pp2}
\end{center}
\end{figure}

Earlier investigations indicated also a considerable sensitivity to
the employed $NN$ interaction model. E. g., the results based on the
Reid soft-core and the Bonn A (r-space version) potentials, respectively,
which were presented by Horowitz et al. in Ref. \cite{HMG}, differ
by almost a factor 2. (Note that such a sensitivity was denied in the
abstract of this paper!) We show here results for the
Paris $NN$ potential \cite{Paris} for which, however,
a somewhat less pronounced variation is observed.
Still its prediction is about 30 \% larger than the one
for OBEPT (cf. the dashed curve in Fig.~\ref{pp2}b). The origin of
this sensitivity can be traced to a node in the $NN$ $^1S_0$
half-off-shell T-matrix occurring at a (off-shell) momentum of around
370 MeV/c. This value is more or less identical with the typical momentum
transfer between the nucleons at pion production threshold (cf. the
comments in the Introduction) and
consequently with the momentum at which the $NN$ half-off-shell
T-matrix is needed for the evaluation of the direct production diagram.
This explains why the contribution of the direct
pion production is relatively small (cf. Fig.~\ref{pp1}). On the
other hand it also means that the magnitude of this contribution
will depend strongly on the specific position of this
node the presence of
which is associated with the transition from the intermediate
attraction to the short-range repulsion in the $NN$ force.

The evaluation of the rescattering- and HME diagrams involves loop
integrations. This means that one averages over the $NN$ half-off-shell
T-matrix. Therefore the dependence of their contributions
on the employed $NN$ models is much less pronounced.

The sensitivity to the $\pi NN$ form factor (applied at the pion
production vertex of the rescattering diagram)
is demonstrated in Fig.~\ref{pp2}a.
We show results for $\Lambda_\pi$ = 800 MeV (used in the present
model) and $\Lambda_\pi$ = 1000 MeV. The latter value is suggested by
a recent study of the $\pi NN$ form factor in the meson-exchange
picture \cite{Jans}. Obviously variations of the cutoff mass within
this range have very little influence on the results.

\begin{figure}[h]
\begin{center}
\epsfig{file=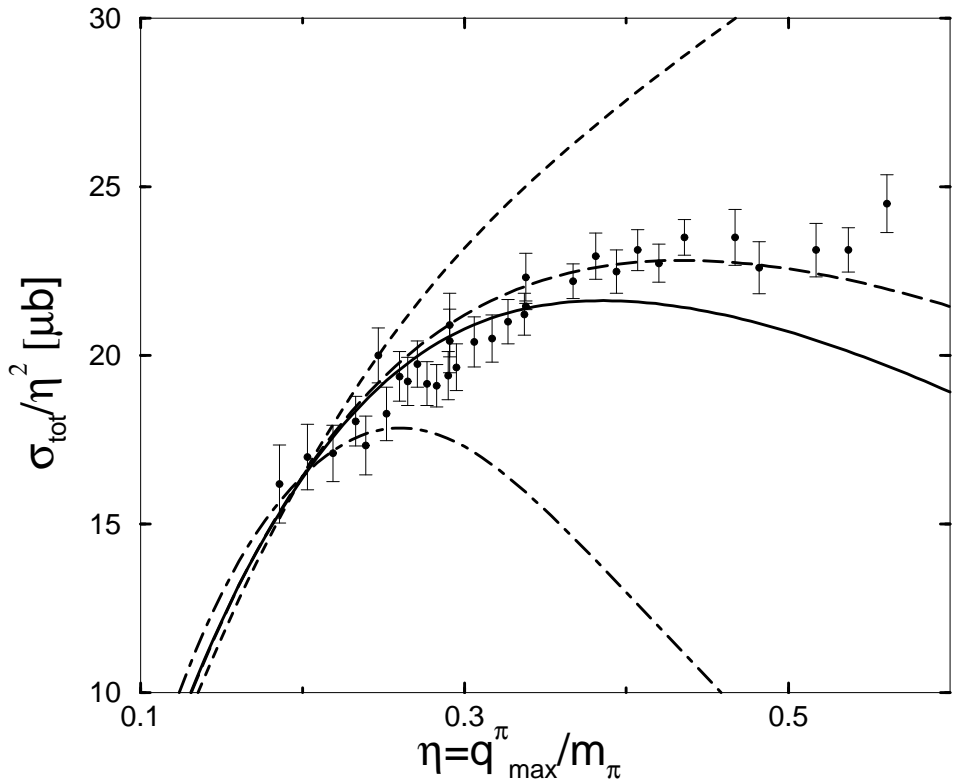, height=12cm}
\caption{Energy dependence of the different contributions to
the reaction $pp \rightarrow pp\pi^0$.
Shown are the direct production (dashed--dotted),
rescattering (long dashed) and heavy meson exchanges (dashed),
each scaled individually to the data at $\eta = 0.2$.
The solid line represents our total result, cf. Fig.~\ref{pp1}.
Note that the cross section is divided by $\eta^2$.
}
\label{enabh}
\end{center}
\end{figure}

A possibility to pin down the contributions of the various pion
production mechanisms
could be provided by a careful study of the energy dependence of the
production cross section. In Fig.~\ref{enabh} we show results for the
contributions of the direct diagram, rescattering and HME individually,
each of them scaled to the data at $\eta = 0.2$. It is evident that
the different production mechanisms lead to rather large differences
in the predicted energy dependence. Unfortunately, for $\eta > 0.4$
where the differences are becoming more pronounced contributions
from ($NN$) p-waves are presumably no longer negligible.
Therefore it is necessary to include these p-waves in the calculations
if one wants to reach reliable conclusions \cite{Han2}.

Another possibility to learn more about the individual production
mechanisms is offered by the study of other $NN \rightarrow NN \pi$
processes ($pn \rightarrow d\pi^0$, etc.) using the same model
(with the same parameters). This is the topic of the next section.

\section{The $pn \rightarrow d \pi^0$ reaction}

\begin{figure}[h]
\begin{center}
\epsfig{file=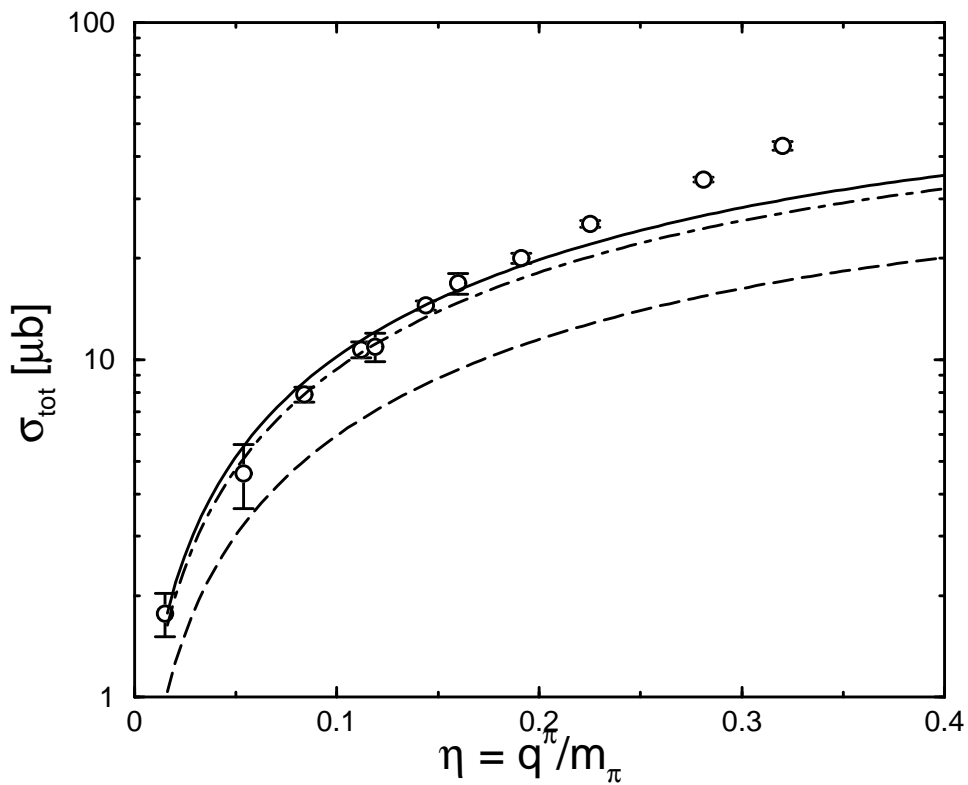, height=12cm}
\caption{Total cross section for the reaction $pn \to d\pi^0$. The
dashed line shows the result for direct production plus isovector
rescattering. Adding isoscalar rescattering yields the dashed--dotted
curve. Including also contributions from heavy meson exchanges
leads to the solid line. The data are from Ref. \protect\cite{Hu}.
}
\label{dpi1}
\end{center}
\end{figure}

Let us now consider the reaction $pn \rightarrow d \pi^0$ close to
threshold. We have calculated the production cross section with the
same model and applying the same parameters (for the pion production
vertex and the heavy meson exchanges)
as for $pp \rightarrow pp\pi^0$. Thus
the results, which are shown in Fig.~\ref{dpi1}, can be considered as
real predictions of our model. The cross section generated by the direct
production mechanism (Fig.~\ref{pm1}a) is extremely small which is due
to the known cancellation between the contributions from the deuteron
s- and d-wave components \cite{KuR,Horr}. Therefore the corresponding curve
cannot be seen in Fig.~\ref{dpi1}.

The bulk of the cross section is provided by isovector s-wave pion
rescattering. It accounts for roughly one half of the experimentally
observed production rate. However, there is also an important
contribution from the isoscalar part of the rescattering process.
Indeed it enhances the cross section by about 50 \% and brings the
result close to the experiments (cf. Fig.~\ref{dpi1}).
Note that this enhancement is entirely due to the fact that
the off-shell properties of the $\pi N$ interaction
are taken into account.
For the static approximation used, e.g., in Ref.~\cite{Horr} the
contribution from the isoscalar channel would be negligible -
like in the corresponding $pp \rightarrow pp\pi^0$ case.

The addition of the contributions from heavy meson exchanges leads only
to a moderate change in the cross section. This is in contradiction
to the results presented in Ref.~\cite{Horr}, where the HME contributions
almost doubled the cross section. We want to mention, however, that
some unjustified assumptions have been made in the aforementioned
calculations, which lead to an overestimation of the effect from
HME by a factor of 3-4, as has been pointed out by Niskanen recently
\cite{nis3}.

Our result starts to deviate from the data at energies around
$\eta =0.25$. We believe that this is due to p-wave contributions which
are missing in our model calculation and which set in at much lower
energies in the reaction $pn \rightarrow d\pi^0$ than in $pp \rightarrow
pp\pi^0$.

The energy dependence of the $np \rightarrow d\pi^0$
cross section near threshold is commonly parameterized by

\begin{equation}
  \sigma _{np} \ = \ {1 \over 2} (\alpha \eta + \beta \eta^3),
\end{equation}

\noindent
where the first coefficient, $\alpha$, is determined by s-wave
pion production. Its experimental value has been given by
Hutcheon et al. to be
$\alpha \ = \ 184 \pm 5 \pm 13 \ \mu b$ \cite{Hu}.
Our model predicts a values of $\alpha \ = 204 \ \mu b$ which
lies just outside of the experimental error bars.

\begin{table}[h]
\label{alpha}
\begin{center}
\begin{tabular}{|l|c|}
\hline
production mechanism & $\alpha \ [\mu b]$ \\
\hline
$\phantom{+}$ direct emission & 0.2 \\
+ isovector rescattering &  120  \\
+ isoscalar rescattering & 189 \\
+ heavy meson exchanges &  204 \\
\hline
\end{tabular}
\end{center}

\caption{Cross section factor $\alpha$, cf. Eq. (3), for the
different production mechanisms.
}
\end{table}

The contribution of the individual production mechanisms are
specified in Table 2.
The HME contribution changes $\alpha$ by only 15 $\mu b$
- a value which is in good agreement with the corresponding results
obtained by Niskanen \cite{nis3}.

\begin{figure}[h]
\begin{center}
\epsfig{file=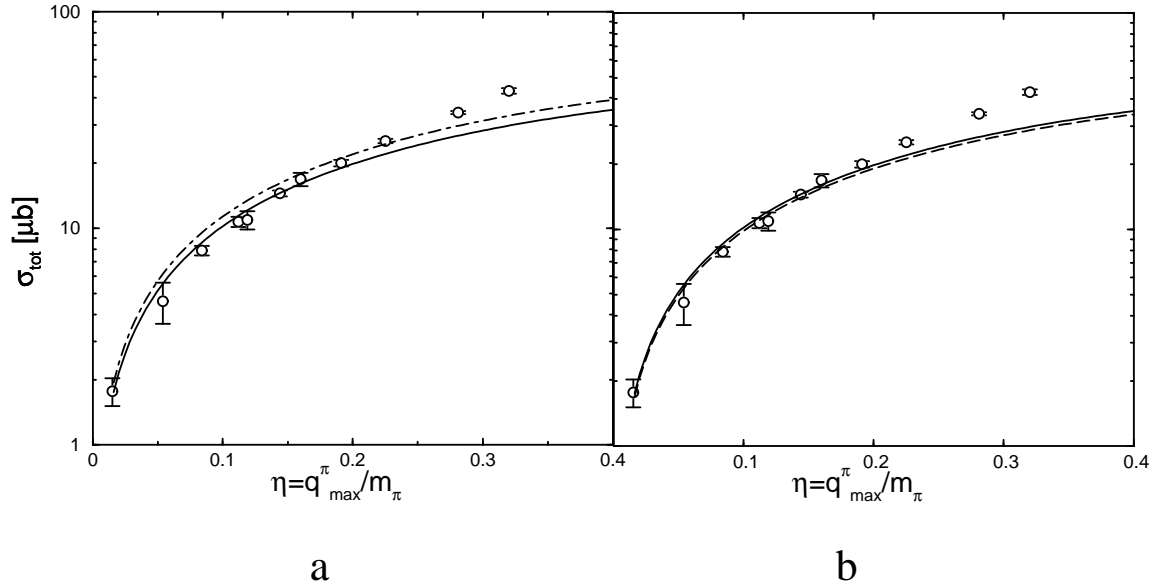, height=8cm}
\caption{Sensitivity of the reaction $pn \to d\pi^0$ to (a) the
$\pi$NN form factor and (b) different nucleon--nucleon interactions.
The solid line is obtained with OBEPT and $\Lambda_\pi$ = 800 MeV.
The dashed--dotted line (in (a)) corresponds to $\Lambda_\pi$ =
1000 MeV. The dashed line (in (b)) is the result for the Paris
$NN$ potential.
}
\label{dp2}
\end{center}
\end{figure}

In Fig.~\ref{dp2} we compare the results obtained with OBEPT with
the ones using the Paris model for the initial- and final state
distortions. Evidently the cross section for $pp \rightarrow pp\pi^0$
is rather insensitive to the used $NN$ interaction. This is in agreement
with the findings reported by Horowitz in Ref. \cite{Horr}. At first
this is surprising because the deuteron wave function of the s-state
(like the $^1S_0$ in the corresponding $pp \rightarrow pp\pi^0$ case)
has also a node around the momentum $q = 370 \ MeV/c$ typical for the
threshold kinematics and its position is again slightly different
for different $NN$ models. However, the contribution of the direct
pion production - which is primarily sensitive to the position of this
node - is very small in case of the reaction $pn \rightarrow d\pi^0$
(cf. the discussion above) and therefore variations of it have a
very small influence on the total production cross section.

The sensitivity of the $pn \rightarrow d\pi^0$ cross section to the
$\pi NN$ form factor is depicted in Fig.~\ref{dp2}a. We see that the
variation in the cross section caused by changing the cutoff mass from
800 to 1000 $MeV$ is somewhat larger than for the reaction
$pp \rightarrow pp\pi^0$ (Fig.~\ref{pp2}a).

\section{Summary}

The hitherto existing investigations on the reaction $pp\rightarrow
pp\pi^0$ have shown that the empirical cross section near threshold
can be reproduced quantitatively. This could be achieved with
contributions from either heavy meson exchanges or from off-shell
pion rescattering so that it seemed that the two production
mechanisms would exclude each other. Our investigations indicate
that they can be combined with each other. Indeed,
within our model contributions from $\pi N$ s-wave rescattering as
well as from heavy meson exchanges are necessary in order to
obtain agreement with the experiments.

The studies indicate
that the production cross section is very sensitive to the short
range component of the $NN$ interaction. First of all this is
reflected in a dependence of the results on the employed $NN$
model. Secondly, it manifests itself in the potentially large
contributions from the HME production mechanism.
In addition there is a strong sensitivity to the off-shell
behavior of the $\pi N$ interaction. Unfortunately, the
uncertainties inherent in either of those properties make it
difficult to disentangle them in a study of the reaction $pp\rightarrow
pp\pi^0$.

One possible way out of this dilemma could lie in a careful analysis
of the energy dependence of the production cross section. As we have
shown, the different production mechanisms lead to rather
pronounced variations. However, here it is necessary to include
also p-waves into the calculations before any reliable conclusions
can be drawn.

Another possibility to learn more about the individual production
mechanisms is offered by the study of other $NN \rightarrow NN \pi$
processes within the same model. For that purpose we looked
at the reaction $pn\rightarrow d\pi^0$. Indeed we found that
this process is much less sensitive to the short-range part of the
$NN$ interaction. There is almost no dependence on the employed
$NN$ model and also contributions from HME play only a minor role.
The bulk of the $pn \rightarrow d\pi^0$ cross section near threshold
is provided by rescattering due to the isovector part of
the s-wave $\pi N$ interaction. However, the contribution
of the isoscalar part is significant so that a comparison with the
data would definitely allow to obtain constraints on the corresponding
off-shell properties.
The isoscalar amplitude of the $\pi N$ model employed in
our study enhances the cross section by about 50 \% and therefore
is responsible for the good agreement of our calculation with the data.
Clearly,
since pion production mechanisms involving the $\Delta$ excitation
(which are so far missing in our model)
could be also important in the reaction $pn \rightarrow
d\pi^0$, as shown by Niskanen, one should be cautious in drawing
quantitative conclusions from this result at present.
\vfill \eject
\end{document}